\documentclass[print,showpacs,twocolumn,superscriptaddress]{revtex4}
\usepackage{amsmath}
\usepackage{graphicx}
\usepackage{mathrsfs}
\usepackage{amssymb}
\usepackage{amsfonts}
\usepackage{amsmath}
\usepackage{graphicx}

\begin{document}
\title{Nuclear spin qubits in a trapped-ion quantum computer}
\author{M. Feng}
\email{mangfeng@wipm.ac.cn} \affiliation{State Key Laboratory of
Magnetic Resonance and Atomic and Molecular Physics, Wuhan Institute
of Physics and Mathematics, Chinese Academy of Sciences, Wuhan
430071, China} \affiliation{Center for Cold Atom Physics, Chinese
Academy of Sciences, Wuhan 430071, China}
\author{Y. Y. Xu,$^{1,2,3}$ F. Zhou}
\affiliation{State Key Laboratory of Magnetic Resonance and Atomic
and Molecular Physics, Wuhan Institute of Physics and Mathematics,
Chinese Academy of Sciences, Wuhan 430071, China}
\affiliation{Center for Cold Atom Physics, Chinese Academy of
Sciences, Wuhan 430071, China} \affiliation{Graduate School of the
Chinese Academy of Sciences, Beijing 100039, China}
\author{D. Suter}
\email{Dieter.Suter@uni-dortmund.de}
\affiliation{Fakult\"{a}t Physik, Technische Universit\"{a}t Dortmund, 44221 Dortmund, Germany}

\begin{abstract}
Physical systems must fulfill a number of conditions to qualify as
useful quantum bits (qubits) for quantum information processing,
including ease of manipulation, long decoherence times, and high
fidelity readout operations. Since these conditions are hard to
satisfy with a single system, it may be necessary to combine
different degrees of freedom. Here we discuss a possible system,
based on electronic and nuclear spin degrees of freedom in trapped
ions. The nuclear spin yields long decoherence times, while the
electronic spin, in a magnetic field gradient, provides efficient
manipulation, and the optical transitions of the ions assure a
selective and efficient initialization and readout.
\end{abstract}


\maketitle

\section{Introduction}

Trapped ions have been recognized as one of the most promising
candidates for future quantum computing (QC) devices for some time
\cite {cz}. This potential has been verified by initial
demonstrations implementing simple quantum algorithms with trapped
ions (see, e.g., \cite{Gulde}) or by entangling eight ultracold ions
in a linear trap \cite {eight}. Nevertheless, a number of difficulties have to be
overcome to scale these devices to larger numbers of qubits. The
proposals for overcoming these obstacles include segmented traps
where ions are transferred between different parts of the trap
\cite{wineland} or by joining many spatially separated traps in a
quantum network \cite{monroe}.

Different groups rely on different ions for their experiments. The NIST group
mainly uses $^{9}$Be$^{+}$ \cite{monroe1}, the groups in Innsbruck, Oxford, Atlanta and Wuhan employ
$^{40}$Ca$^{+}$ \cite{hafner,lucas,chapman,wuhan}, and the group in Maryland works with
$^{111}$Cd$^{+}$ \cite{monroe2} and $^{171}$Yb$^{+}$ \cite{monroe3}.
There are also some proposals and
experimental reports using $^{171}$Yb$^{+}$ \cite{wund,wund2}, $^{43}$Ca$^{+}$
\cite{blatt,blatt2}, $^{138}$Ba$^{+}$ \cite{blinov}, and $^{88}$Sr$^{+}$ \cite {chuang}.
In all these cases, however, the qubits are encoded in two
electronic states and the exchange of quantum information between ions occurs via
quantized vibrational modes.
The main conditions for these schemes to work include long lifetimes of the computational states
and well resolved vibrational modes.
Recent experiments have demonstrated that quantum information can be conserved
almost perfectly even when the ions are moved \cite{lieb}.

In the context of solid-state QC, several proposals have been put
forward that suggest the use of nuclear spins for storing quantum
information \cite {roadmap}. Nuclear spins are very well isolated
from other degrees of freedom and therefore promise long storage
times for quantum information. On the other hand, the weak
interaction with their environment also makes initialization and
manipulation of nuclear spins difficult and leads to low detection
sensitivity. It may therefore be necessary to combine nuclear spins
with other degrees of freedom to combine long storage times with
efficient manipulation and detection
\cite{kane,dieter,Harneit,twam,du}. Of these proposals, so far only
one system has been demonstrated experimentally, which combines the
electronic spin of the diamond NV center with a neighboring $^{13}$C
nuclear spin \cite{lukin,Neumann}.

In this paper, we propose to combine nuclear spins with trapped
ions. The goal is to design a system that combines long decoherence
times with efficient manipulation and detection. The trapped-ion
system that we consider is closely related to the proposal by
Mintert {\it et al} \cite{mintert}, which relies on a magnetic field
gradient to address different ions and to induce couplings
between the ions that do not rely on the motional degrees of
freedom. This scheme can be used with different ions, such as
$^{43}$Ca$^{+}$, whose nuclear spin is $I=7/2$,  $^{135}$Ba$^{+}$ and $^{137}$Ba$^{+}$ with I=3/2 as
well as $^{171}$Yb$^{+}$ with $I=1/2$. For specific discussions, we will
refer to the $^{43}$Ca$^{+}$ system, but all our results are also applicable to the other three candidates.

The hyperfine splitting of the $S_{1/2}$ ground state of
$^{43}$Ca$^{+}$ has been precisely measured \cite{arbes,blatt1}.
Since our scheme works best when nuclear and electronic spins can be
clearly distinguished, we explicitly consider the case of a strong
magnetic field (i.e., the Paschen-Back regime).

The paper is structured as follows: In the following section, we
first consider a single trapped ion and show how the quantum
information can be encoded in the nuclear spin degree of freedom and
transferred to the electronic spin. Section III introduces two-qubit
operations and in section IV, we discuss the operation of the full
quantum register, including addressing of individual qubits. Section
V presents a short summary and conclusions.

\section{Single $^{43}$Ca$^{+}$ ion}
\label{s.SingleIon}

We first consider a single trapped $^{43}$Ca$^{+}$ ion.
In the Paschen-Back regime, we can write the Hamiltonian as%
\begin{equation}
H_{0}=\Omega_{S}S_{z}-\Omega_{I}I_{z}+A S_{z}I_{z},
\label{e.1}
\end{equation}
where we use units of $\hbar=1$. $S_{z}$ and $I_{z}$ are the spin
operators of the electron spin ($S=1/2$) and the nuclear spin
($I=7/2$). $\Omega_{S}=g_{S}\mu_{B}B$ and $\Omega_{I}=g_{n}\mu_{B}B$
are the Larmor frequencies of the electron and nuclear spins, and
$A=-806.4$ MHz \cite{blatt1} is the hyperfine coupling constant.
We include here only the high-field truncated part of the hyperfine coupling.
In the absence of external radiation, the vibrational mode is not
coupled to the spin degrees of freedom and was therefore not
included in the Hamiltonian of Eq. (1).

\begin{figure}[htb]
\begin{center}
\includegraphics[width= 1.0\columnwidth]{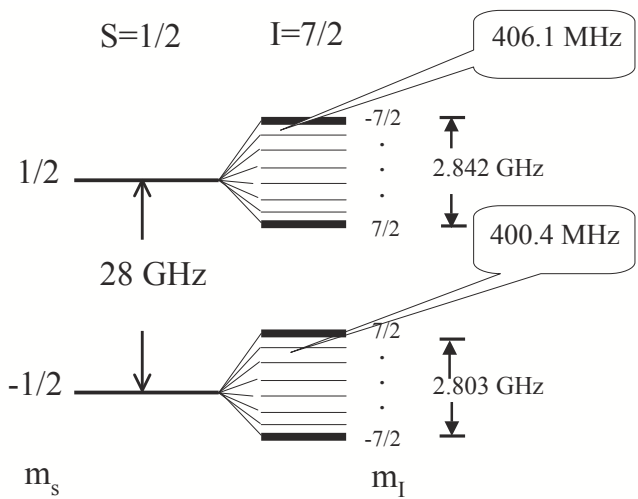}
\end{center}
\caption{Angular momentum states in the ground state of a
$^{43}$Ca$^{+}$ ion in a magnetic field $B=1$ T (i.e. in the
Paschen-Back regime). The four levels drawn in bold are used to
define two qubits.} \label{f.levels}
\end{figure}

Figure \ref{f.levels} shows the resulting energy level structure in a magnetic field of $B=1$ T.
The electron Zeeman interaction causes a splitting of 28 GHz.
Each of these electronic states is split into 8 nuclear spin sublevels by the hyperfine interaction.
Due to the nuclear Zeeman interaction, the splittings within the two multiplets differ by 5.7 MHz.

\begin{figure}[htb]
\begin{center}
\includegraphics[width= 1.0\columnwidth]{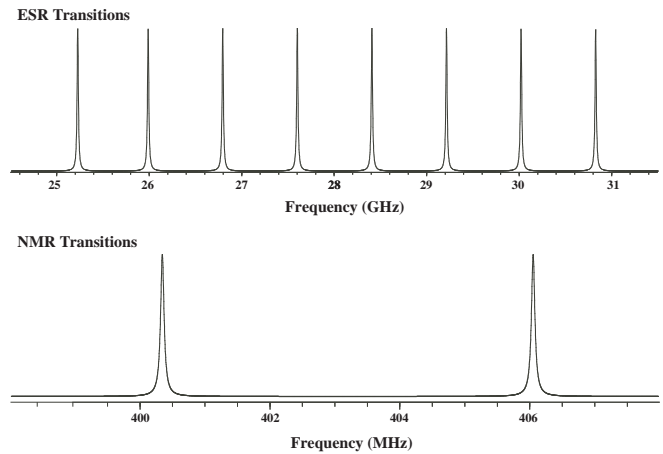}
\end{center}
\caption{Magnetic dipole transitions in the $^{43}$Ca$^{+}$ spin
system. The upper trace shows the transitions that correspond to a
change of the electron spin. They fall into the microwave range,
close to 28 GHz, and are split by the hyperfine interaction into 8
transitions, each corresponding to a single nuclear spin state. The
lower trace shows the transitions of the nuclear spin, which fall
into the radio frequency range. The lower frequency line corresponds
to the $m_S$=-1/2 state of the electron spin, and the higher
frequency transition to the $m_S$=+1/2 state. } \label{f.spec}
\end{figure}

Transitions between these spin states can be induced by the magnetic
dipole interaction. Changes of the electron spin correspond to
frequencies in the microwave range, centered around 28 GHz, as shown
in the upper part of Fig.\,\ref{f.spec}. The lower part shows the
transitions corresponding to changes of the nuclear spin state.
Their frequencies correspond to $|A/2|\pm\Omega_I$.

While this system could in principle store 4 qubits (i.e., three in
the nuclear spin and one in the electronic spin), we will consider
here only two qubits. The first qubit corresponds to the electron
spin and the second qubit will be encoded in the $m_I = \pm 7/2$
states of the nuclear spin.

To apply gate operations to these qubits, we use resonant microwave radiation,
\begin{equation}
H_{1}=H_{0} + 2 \cos(\Omega_{\mathrm{mw}} t) [\omega_S S_{x} +  \omega_I I_{x}].
\label{e.2}
\end{equation}
Here, $H_{0}$ is the system Hamiltonian of Eq. (\ref{e.1}),
$\omega_S$ and $\omega_I$ are the Rabi frequencies of the electronic
and nuclear spins and $\Omega_{\mathrm{mw}}$ the frequency of the
applied field.

In the following, we will assume that the microwave field is weak enough that it interacts
predominantly with a single transition.
This is fulfilled if the Rabi frequencies are small compared to the splittings between the resonance lines,
$$
\omega_S \ll |A| \quad \mathrm{and}\quad \omega_I \ll |\Omega_I| \,.
$$
The electron spin qubit can then be manipulated by microwave
radiation, if the microwave field is resonant with one of the two
transitions at 25.2 and 30.8 GHz, which correspond to the nuclear
spin being in the $m_I = 7/2$ and -7/2 states, respectively. A
microwave pulse at one of these frequencies thus corresponds to a
single qubit rotation of the $S$-spin, conditioned on the $I$-spin
being in the corresponding state. Unconditional single-qubit
operations on the $S$-spin can be accomplished as a sequence of two
(or two simultaneous) conditional operations.

Applying single qubit gates to the nuclear spin qubit is less straightforward:
In general, a resonant RF pulse takes the nuclear spin out of the state space
associated with the nuclear spin qubit ($m_I = \pm 7/2$).
An important exception is the operation $P_{I}=e^{-i\pi I_{x}}$, which
simply interchanges the two states $|7/2\rangle$ and $|-7/2\rangle$ :
\begin{equation}
e^{-i\pi I_{x}}I_{z}e^{i\pi I_{x}}=-I_{z} \, . \label{3}%
\end{equation}

If we use resonant RF irradiation, the exchange becomes again
conditioned on the state of the electronic qubit. As an example, a
RF pulse at $\Omega_{\mathrm{mw}}$ = 406.1 MHz exchanges $\left\vert
7/2\right\rangle _{I}\left\vert 1/2\right\rangle _{S} \iff\left\vert
-7/2\right\rangle _{I}\left\vert 1/2\right\rangle _{S}$, while the
$\left\vert 7/2\right\rangle _{I}\left\vert -1/2\right\rangle _{S}$
and $\left\vert -7/2\right\rangle _{I}\left\vert -1/2\right\rangle
_{S}$ states remain invariant. This corresponds to a CNOT operation,
which we call CNOT$_{\mathrm{SI}}$: the first index refers to the
control qubit and the second to the target qubit. Similarly, a
resonant $\pi$ pulse applied to the ESR transition at 25.2 GHz
corresponds to a CNOT$_{\mathrm{IS}}$ operation.

We use these operations to implement a SWAP operation between
electronic and nuclear qubits:
\begin{eqnarray*}
\mathrm{SWAP} & = & \mathrm{CNOT_{IS}}\,\mathrm{CNOT_{SI}}\,\mathrm{CNOT_{IS}}\, \\
& = & \mathrm{CNOT_{SI}}\,\mathrm{CNOT_{IS}}\,\mathrm{CNOT_{SI}}\,.
\end{eqnarray*}

The SWAP operation can now also be used to implement arbitrary single-qubit operations $U_I$ on the nuclear
spin qubit:
$$
U_I = \mathrm{SWAP} \, \, U_S  \, \,  \mathrm{SWAP} \, ,
$$
where $U_S$ is the single-qubit operation acting on the electron
spin qubit. The system allows thus to implement arbitrary one- and
two-qubit operations.

\section{A pair of qubits}

\subsection{System and Hamiltonian}

To implement a quantum register on the basis of the $^{43}$Ca$^{+}$
system, we need a string of trapped ions. We will use the $|\pm7/2
\rangle$ states of the nuclear spins as the qubits, while the
electron spins are used to couple the nuclear spin qubits to each
other.

When a magnetic field gradient $B(z)=B_{0}\,+\,b \, \delta z$ is applied
with $b=\partial B/\partial z$, the Hamiltonian of the system is
\begin{eqnarray}
H_2 & = & \sum_i \Omega_S^i S_z^i -  \sum_i \Omega_I^i I_z^i + A\sum_i S_z^i I_z^i \nonumber \\
& & - \sum_{i<j}J_{ij}S_z^i S_z^j \, ,
\label{e.Ham2}
\end{eqnarray}
where the first and the second terms represent the Zeeman energies of the
electronic and the nuclear spins, respectively, and the third term is the hyperfine interaction.
The Larmor frequencies $\Omega_S^i$ and $\Omega_I^i$ depend on the magnetic field gradient $b$
and the position $z_i$ of the ions as
$$
\Omega_S^i = g \mu_{B}(B_0 + b \, z_i)
$$
and correspondingly for the nuclear spins.

The fourth term in Eq. (\ref{e.Ham2}) represents a coupling between
the different ions, which takes the form of an Ising interaction.
This effective interaction is generated by the magnetic field
gradient. The coupling constants $J_{ij}$ are proportional to the
square of the ratio of the magnetic field gradient to the center of mass
frequency \cite{wund,deng,twam1}. It determines the gate operation
time $T = \pi / J_{12}$ for two-qubit operations. In Table
\ref{tab.gs}, we list some possible combinations of trap parameters
together with the corresponding gate operation times. For
simplicity, we have omitted the corresponding interaction between
the nuclear spins, which is very small compared to all other terms.

\begin{widetext}
\begin{figure}[htb]
\begin{center}
\includegraphics[width= 1.6\columnwidth]{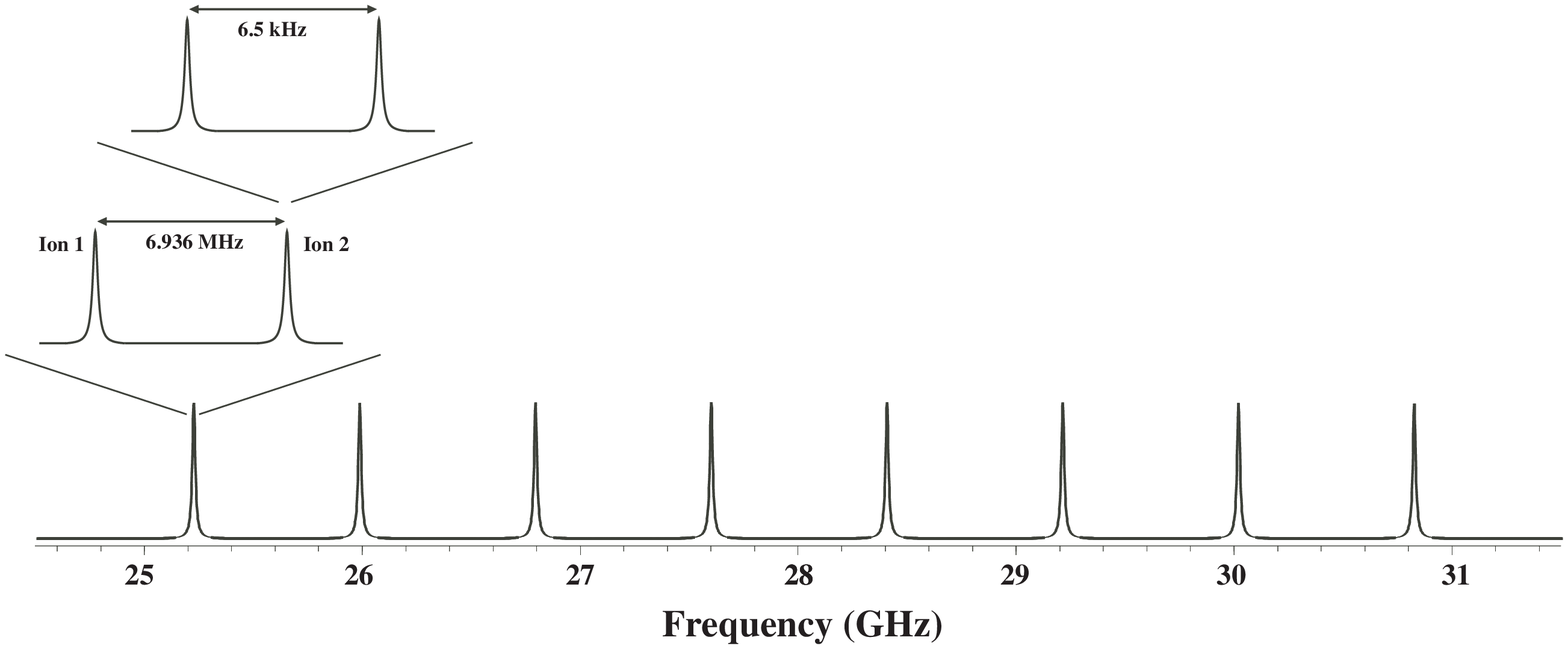}
\end{center}
\caption{ESR spectrum of two ions in a trap. The lowest trace shows
the full spectrum, the middle trace shows the lowest frequency line
on an expanded scale. The splitting is due to the magnetic field
gradient. The uppermost trace shows the splitting due to the
J-coupling between the ions. } \label{f.2Ions}
\end{figure}
\end{widetext}

We first consider the case of two ions in the trap. For the
parameters corresponding to the first line in Table \ref{tab.gs},
the spectrum of the ion pair remains essentially the same as that of
the single ion, as shown on the bottom of Fig.\,\ref{f.2Ions}. The
separation of the resonance lines due to the magnetic field gradient
becomes apparent on an expanded frequency scale, as shown in the
central trace of Fig.\,\ref{f.2Ions}. The splitting due to the
gradient-induced J-coupling between the ions is then another 3
orders of magnitude smaller, as shown in the top trace.

\begin{table}[htdp]
\caption{ $\mathrm{CNOT_{S_1S_2}}$ gating time $T$ for a pair of
trapped $^{43}$Ca$^{+}$ in a 1 T magnetic field for different center-of-mass trap frequencies
$\nu_{1}$, magnetic field gradients $b$, and distances $\Delta
z_{\min}$. }
\begin{center}
\begin{tabular}[c] {|l|l|l|l|l|} \hline
$\nu_{1}/2\pi$ $($MHz$)$ & $b$ $($T/m$)$ & $\Delta z_{\min}$
$(\mu$m$)$ & $J_{12}$ $($kHz$)$ & $T$ $($ms$)$\\\hline $1.0$ & $450$
& $5.5$ & $3.25$ &  $0.24$\\\hline $1.0$ & $230$ & $5.5$ & $0.85$ &
$0.92$\\\hline $1.0$ & $50$ & $5.5$ & $0.40$ &  $1.75$\\\hline $0.8$
& $340$ & $6.3$ & $2.90$ & $0.27$\\\hline $0.8$ & $160$ & $6.3$ &
$0.65$ & $1.21$\\\hline $0.8$ & $35$ & $6.3$ & $0.30$ &
$2.62$\\\hline
\end{tabular}\end{center}
\label{tab.gs}
\end{table}

\subsection{Two-qubit operation}

A CNOT operation between the two qubits can be achieved via the following three steps:

(1) The information encoded in nuclear spins is transferred to the electronic
spins by the gates SWAP$_\mathrm{^{_{I_{1}S_{1}}}}$ and SWAP$_\mathrm{^{_{I_{2}S_{2}}}}$.

(2) CNOT$_\mathrm{S_{1}S_{2}}$ is applied to the electronic spins.

(3) Step (1) is repeated to transfer the information back to the nuclear spins.

Looking in more detail at the second step, we note that the
CNOT$_\mathrm{S_{1}S_{2}}$ operation does not involve the nuclear
spin. We may thus disregard the nuclear spin degrees of freedom
during this step. The problem reduces then to the conventional
system of two spins 1/2, which are described by an effective
Hamiltonian
\begin{equation}
H_{3}=\Omega_{1} S_{z}^{1} + \Omega_{2} S_{z}^{2} - J_{12}S_{z}^{1}S_{z}^{2}.\label{5}%
\end{equation}
Here, the Larmor frequencies $\Omega_{1,2}$ of the two electron spins include a contribution from the
hyperfine coupling,
$$
\Omega_1 = \Omega_S^1 + m_{I_1} A \, ,
$$
and analogously for the second ion. Here, $m_{I_1}=\pm7/2$ denotes
the nuclear spin state. This Hamiltonian is identical to that of a
weakly coupled pair of spins 1/2 in liquid-state NMR. The CNOT
operation can therefore be carried out in complete analogy, either
by a selective pulse on one of the resonance lines corresponding to
the transition between the states (1,0) and (0,1) \cite{KumarSelP},
or by two $\pi/2$ pulses applied to the first qubit, separated by a
free precession period of duration $\pi/J_{12}$ under the coupling
Hamiltonian \cite{NielsenChuang,StolzeSuter}.

\subsection{Selective SWAP operations}

The SWAP operations required in steps (1) and (3) were
discussed in section \ref{s.SingleIon}. However, since the quantum
register now contains multiple ions, we need more specific control
over the radiofrequency and microwave fields. Since the overall
duration of any algorithm is determined mostly by the relatively
weak J-couplings, it is not necessary to make the Rabi frequencies
very high.
A useful Rabi frequency for the microwave pulses would
then be of the order of 1 MHz, which can easily be achieved with
very small microwave power.
As an example, the devices described in \cite{MicroRes} can generate 1 MHz Rabi frequencies
with 6 $\mu$W of microwave power.
Being much smaller than the hyperfine coupling constant, the microwave will excite only the transition
corresponding to the desired nuclear spin state and only the
targeted ion, while the neighboring ions will not be affected. Such
a pulse will, however, excite both lines of the doublet due to the
J-coupling.

\subsection{Initialization and readout}

Apart from the gate operations discussed so far, initialization and
readout are essential steps for the implementation of a quantum
information processor. In this respect, our proposal is almost
identical to that of conventional ion trap quantum computers
\cite{bw}: Initialization is achieved by optical pumping, while
readout is achieved by electronic shelving amplification \cite{bz}.
Since the addressing of the ions as well as the interactions between
them are magnetic, there is no need for cooling the ions to their
ground state \cite{wund}, but only into the Lamb-Dicke limit.

The main difference arises from the fact that the qubits are encoded
in the nuclear spin degrees of freedom: After the initialization, it
is thus necessary to SWAP the initialized state into the nuclear
spin. The same process is required in reverse for readout: The
resulting state must be SWAPed from the nuclear spin degree of
freedom to the electronic degree of freedom, where readout occurs.

\section{Quantum register}

\subsection{Addressing qubits}

For an actual QIP implementation, more than two ions are required.
In terms of trap dynamics, it should be possible to store more than ten ions in such a trap.
Addressing of the ions is straightforward in the scheme described here, as long as the Larmor frequencies
of the electronic spins do not overlap.
This requires
\begin{equation}
g \mu_B b \Delta z N < A \, ,
\end{equation}
where $N$ is the number of ions in the string.

Addressing the individual qubits can then be achieved by keeping the microwave field strength well below
the difference of the Larmor frequencies of two neighboring ions,
\begin{equation}
\omega_1 < g \, \mu_B \, b \, \Delta z  \, .
\end{equation}
This condition can be easily met without increasing the total duration of any quantum algorithm.

So far, we have only considered the electronic spins. But as we have
discussed in the previous section, actual gate operations require
resonant excitation of the electronic as well as the nuclear spin.
The magnetic field gradient also separates the nuclear Larmor
frequencies by
\begin{equation}
\Delta \Omega_I = - 18 b \Delta z  \, ,
\end{equation}
with $\Omega_I $ in MHz, $b$ in T/m, and $ \Delta z$ in m.
Compared to the electron spin resonance frequencies, this splitting
is reduced by approximately 4 orders of magnitude, which amounts to
$\approx 700$ Hz for the parameters used above. Addressing the
nuclear spins individually would thus increase the duration of a
quantum algorithm significantly. However, this can be easily avoided
if we notice that only $\pi$ pulses are applied to the nuclear spins
and that all relevant operations discussed in sections II and III
involve two SWAP operations between the electron and nuclear spins
of the ion and thus an even number of $\pi$ pulses applied to the
nuclear spins.
We can therefore apply the RF pulses nonselectively,
i.e. to all nuclear spins in the quantum register, using a strong rf
field and short pulse durations.
The individual $\pi$ pulses will then affect all
\emph{nuclear} spins in the trap; however, for those ions that are
not addressed by the frequency-selective pulses applied to the
\emph{electronic} spins, the operation of the  $\pi$ pulse is
reversed by the next $\pi$ pulse and results in a NOP for the
inactive qubits.

\subsection{Refocusing operations}

In a string of ions, the induced coupling varies with the position
of the ions in the trap \cite{wund, mintert}. Couplings exist
between all pairs of ions in the trap, not only between nearest
neighbors. Quantum algorithms therefore have to take the full
network of couplings into account. This operational form of the
couplings,  $S^i_z S^j_z$, provides a close analogy to liquid-state
NMR, where undesired couplings are usually eliminated by refocusing
techniques (see, e.g.,
\cite{NielsenChuang,StolzeSuter,suterJCP,Vandersypen,Pravia,Schulte}).

In the present case, such refocusing operations are more straightforward than in the case of liquid state NMR.
If we want to implement a two-qubit operation between ions $i$ and $j$, we want to have one active coupling
between these two ions, while all other couplings should be refocused.
Since all terms in the relevant Hamiltonian in Eq. (\ref{e.Ham2}) commute with each other,
it is sufficient to consider the terms
\begin{equation}
H_4 =  \Omega_S^i S_z^i + \Omega_S^j S_z^j  - J_{ij}S_z^i S_z^j  - J_{ik}S_z^i S_z^k \, ,
\label{e.Ham3}
\end{equation}
where the index $k$ represents any of the passive spins.
The evolution of the two active spins $i$ and $j$ under the coupling $J_{ij}$ is now achieved by
letting the system evolve under this Hamiltonian for a duration $\tau = \pi/(4J_{ij})$,
applying $\pi$ pulses to spins $i$ and $j$, and letting them evolve for another period $\tau$,
and applying another pair of $\pi$ pulses.
The resulting overall evolution is then
\begin{equation}
U_2 = e^{-i H_4 (\pi/4J_{ij})} e^{-i \pi (S_y^i +S_y^j)} e^{-i H_4
(\pi/4J_{ij})} e^{-i \pi (S_y^i +S_y^j)} \, .
\end{equation}
Using
\begin{eqnarray}
e^{-i \pi (S_y^i +S_y^j)} H_4  e^{-i \pi (S_y^i +S_y^j)} = \nonumber \\
- \Omega_S^i S_z^i - \Omega_S^j S_z^j  - J_{ij}S_z^i S_z^j  + J_{ik}S_z^i S_z^k \, ,
\end{eqnarray}
we can simplify this to the desired evolution
\begin{equation}
U_2 = e^{-i \frac{\pi}{2} S_z^i S_z^j}  \, .
\end{equation}

\subsection{Decoherence}

Like in other implementations of quantum information processing, the operations must be completed
in a time short compared to the decoherence time.
We consider here two contributions to the decoherence:
(i) dephasing due to magnetic field instability and
(ii) heating of the motional degrees of freedom.

The evolution of the electronic spins in the magnetic field changes the phase of the qubits by
$$
\Delta \varphi_i = \Omega_i \tau \, .
$$
Instabilities of the magnetic field therefore affect the phase of the overall quantum state and must be reduced to
$$
\Delta B \ll \frac{1}{g \, \mu_B \,\tau} \,.
$$
Since the gate operation time $\tau$ is of the order of a few ms,
this requires a stability of the order of $10^{-8}$ T.
Clearly, this is a challenging condition; however, the corresponding magnet technology is well
established in the field of high-resolution NMR, where field stability of better than $10^{-10}$ is required.
The requirements can also be reduced, by
working at lower magnetic field, by refocusing techniques, or by
working in decoherence-free subspaces \cite{lidar}.

Additional contributions to the decoherence of trapped ions are due
to mobile charges in the trap and to motional heating. In the
proposed trap, where the information is encoded in the spin degrees
of freedom, such contributions should be considerably smaller than
in conventional ion traps. Moreover, the couplings between different
electron spins are due to the Coulomb interaction of the trapped
ions under the magnetic field gradient, which only virtually excites
the motional degrees of freedom. This also implies that motional
heating during gate operation is suppressed, as discussed by
Wunderlich and coworkers \cite{mintert,wund}.

\subsection{Experimental feasibility}

A main point of our proposal is that we use only spin degrees of freedom.
This leads to a considerable simplification of the level diagram compared to the case where
electronic degrees of freedom contribute to the relevant Hilbert space spanned by $S_{1/2}$ and $D_{5/2}$ \cite {blatt2}.
The possibility of addressing trapped ions using a magnetic field gradient and resonant
microwave radiation was recently demonstrated using a magnetic field gradient weaker than 1 T/m \cite{wund2,chuang}.
Although such a magnetic field gradient is not big enough
to generate efficient CNOT gates, this should become possible by an optimization of the experimental parameters.
Stronger magnetic field gradients (up to 8000 T/m) can be generated, as discussed by Mintert and Wunderlich \cite{mintert}.

A larger field gradient would provide stronger couplings and
therefore faster two-qubit gate operations. However, the gradient
also couples the electron spin to the vibrational motion and
thereby, if it is too big, reduces the fidelity of the gate
operations. An upper limit to the useable gradient strength is given
by the condition $\epsilon \ll 1$ \cite {mintert}, where the
parameter
$$
\epsilon = \frac{g \mu_B \frac{\partial B}{\partial z}\delta z}{\nu_1}
= \frac{g \mu_B b}{\sqrt{2 N m \nu_1^3}}
$$
in units of $\hbar=1$ is a generalized Lamb-Dicke parameter, with
$m$ the mass of the ion, $N$ the number of the confined ions, and
$\nu_1$ the fundamental vibrational frequency. For our system, where
$\nu_{1}=2\pi$ MHz, $N=2$, $m=43\times 1.67\times 10^{-27}$ kg the
upper limit for the magnetic field gradient becomes $b \le 455$ T/m.
Limitations on the magnetic field gradient for stable trapping of ions are also discussed by Cronin et al. \cite {Schm}.

\section{Discussion and Conclusion}

In this paper, we have proposed a scheme for encoding quantum
information in nuclear spin states of trapped ions. $^{43}$Ca$^{+}$
has also been used in Ref. \cite{blatt2}, where the authors encoded
the information in field-insensitive hyperfine states. Since we work
in the Paschen-Back regime, there are no field-insensitive states.
It is therefore necessary to use stable magnetic fields and,
possibly, shield against undesired external fields. However, the
requirements on the stability of the magnetic field remain
significantly less strict than in the case of liquid-state NMR,
where the field has to be kept stable on a scale of $\approx
10^{-10}$. We conclude that this should not be a significant
obstacle.

The proposed scheme is also related to an earlier proposal for QC with doped fullerenes \cite{dieter,Harneit,du}.
Both use nuclear and electronic spin degrees of freedom and work in the Paschen-Back regime.
In terms of requirements for field stability both systems should be comparable
and the gate operations rely in both cases on resonant microwave pulses.
The main advantage of the trapped-ion system over the fullerene scheme would be the relatively easy readout,
which remains a significant challenge for the fullerene system.
While the solid-state systems may be easier to scale up, our trapped-ion system may also be made
scalable by approaches based on segmented traps \cite{wineland} or multiple connected traps \cite{monroe}.
A third possibility might be the use of microtraps for single ions, which is also compatible with
field-gradient induced couplings between the ions \cite{twam1}.

Throughout this paper, we have relied on field-gradient induced couplings between the ions to implement
multi-qubit gate operations. This is not the only possibility, however, because the use of nuclear
spin degrees of freedom for shelving the quantum
information is also compatible with the more conventional Cirac-Zoller method \cite {cz} or the
S{\o}rensen-M{\o}lmer scheme \cite {sm} which does not require extensive cooling of the ions.

In summary, we have investigated the possibility to carry out QC
using nuclear-spin qubits in $^{43}$Ca$^{+}$ ions. Using the nuclear
spin should result in long decoherence times and working in the
Paschen-Back regime results in a straightforward distinction of
nuclear and electronic spin degrees of freedom. All the required
gate operations can be achieved with microwave pulses, in which
individual addressing of the qubits is achieved by
frequency-selection. If a few-qubit quantum computer on the basis of
this scheme can be implemented, it could serve as a test-bed not
only for larger trapped-ion quantum computers, but also for similar
schemes based on condensed-matter systems, such as endohedral
fullerenes \cite{dieter,Harneit,du} or NV centers in diamond
\cite{lukin,Neumann}.

\section*{ACKNOWLEDGMENTS}

This work is partly supported by NNSF of China under Grant No.
104774163, by the NFRP of China under Grant No. 2006CB921203,
and by the Robert Bosch Stiftung.


\begin{thebibliography}{99}

\bibitem {cz} J. I. Cirac and P. Zoller, Phys. Rev. Lett. \textbf{74}, 4091 (1995).

\bibitem{Gulde} S. Gulde, M. Riebe, G. P. T. Lancaster, C. Becher, J. Eschner, H.
Haeffner, F. Schmidt-Kaler, I. L. Chuang, and R. Blatt, Nature
(London) \textbf{421}, 48 (2003).

\bibitem{eight} H. H\"{a}ffner,
W. H\"{a}nsel, C. F. Roos, J. Benhelm, D. Chek-al-kar, M. Chwalla, T. K\"{o}rber,
U. D. Rapol, M. Riebe, P. O. Schmidt, C. Becher, O. G\"{u}hne, M. D\"{u}r and R. Blatt,
Nature (London) \textbf{438}, 643 (2005).

\bibitem{wineland} D. Kielpinski, C. Monroe and D. J. Wineland,
Nature (London) \textbf{417}, 709 (2002).

\bibitem{monroe} D. L. Moehring, P. Maunz, S. Olmschenk, K. C. Younge, D. N. Matsukevich,
L. M. Duan and C. Monroe, Nature (London) \textbf{449}, 68 (2007).

\bibitem{monroe1} C. Monroe, D. M. Meekhof, B. E. King, W. M. Itano, and D. J. Wineland,
Phys. Rev. Lett. \textbf{75}, 4714 (1995).

\bibitem{hafner} M. Riebe, H. H\"{a}ffner, C. F. Roos, W. H\"{a}nsel, J. Benhelm,
G. P. T. Lancaster, T. W. K\"{o}rber, C. Becher, F. Schmidt-Kaler, D. F. V. James,
and R. Blatt, Nature (London) \textbf{429}, 734 (2004).

\bibitem{lucas} M. J. McDonnell, J. -P. Stacey, S. C. Webster, J. P. Home, A. Ramos,
D. M. Lucas, D. N. Stacey and A. M. Steane, Phys. Rev. Lett. \textbf{93}, 153601 (2004).

\bibitem{chapman} A. V. Steele, L. R. Churchill, P. F. Griffin, and M. S. Chapman, Phys. Rev. A
\textbf{75}, 053404 (2007)

\bibitem{wuhan} H. L. Shu, B. Guo, H. Guan, Q. Liu, X. R. Huang and K. L. Gao,
Chin. Phys. Lett. \textbf{24}, 1217 (2007).

\bibitem{monroe2} B. B. Blinov, D. L. Moehring, L. M. Duan, and C. Monroe, Nature
(London) \textbf{428}, 153 (2004),

\bibitem{monroe3} S. Olmschenk, K. C. Younge, D. L. Moehring, D. N. Matsukevich, P. Maunz, and C. Monroe,
Phys. Rev. A \textbf{76}, 052314 (2007).

\bibitem{wund} C. Wunderlich, C. Balzer, T. Hannemann, F. Mintert, W. Neuhauser,
D. Rei$\beta$ and P. E. Toschek, J. Phys. B \textbf{36}, 1063 (2003).

\bibitem {wund2} M. Johanning, A. Braun, N. Timoney, V. Elman, W. Neuhauser and
C. Wunderlich, \textbf{102}, 073004 (2009).

\bibitem{blatt} C. F. Roos, M. Chwalla, K. Kim, M. Riebe and R. Blatt, Nature
(London) \textbf{443}, 316 (2006).

\bibitem{blatt2} J. Benhelm, G. Kirchmair, C. F. Roos and R. Blatt, Phys. Rev. A \textbf{77}, 062306 (2008).

\bibitem{blinov} N. Kurz, M. R. Dietrich, Gang Shu, R. Bowler, J. Salacka, V. Mirgon, and B. B. Blinov, Phys.
Rev. A \textbf{77}, 060501 (2008).

\bibitem{chuang} S. X. Wang, J. Labaziewicz, Y. Ge, R. Shewmon, and I. L. Chuang, eprint, arXiv: 0811.2422.

\bibitem{lieb} D. Leibfried, B. DeMarco, V. Meyer, M. Rowe, A. Ben-Kish, M. Barrett,
J. Britton, J. Hughes, W. M. Itano, B. M. Jelenkovic, C. Langer, D. Lucas, T. Rosenband
and D. J. Wineland, J. Phys. B \textbf{36}, 599 (2003).

\bibitem{roadmap} For example, {\it A Quantum Information Science and Technology Roadmap} for solid-state systems at
website http://qist.lanl.gov/pdfs/ss.pdf

\bibitem{kane} B. E. Kane, Nature (London) \textbf{393}, 133 (1998).

\bibitem{dieter} D. Suter and K. Lim, Phys. Rev. A \textbf{65}, 052309 (2002).

\bibitem{Harneit} W. Harneit, C. Meyer, A. Weidinger, D. Suter, and J. Twamley, Phys. Stat. Sol. (b) \textbf{233} 453 (2002).

\bibitem{twam} J. Twamley, Phys. Rev. A \textbf{67},  052318 (2003).

\bibitem{du} C. Ju, D. Suter, and J. Du, Phys. Rev. A \textbf{75}, 012318 (2007).

\bibitem{lukin} M. V. G. Dutt, L. Childress, L. Jiang, E. Togan, J. Maze, F. Jelezko,
A. S. Zibrov, P. R. Hemmer and M. D. Lukin, Science \textbf{316}, 1312 (2007).

\bibitem{Neumann} P. Neumann, N. Mizuochi, F. Rempp, P. Hemmer, H. Watanabe, S. Yamasaki, V. Jacques,
T. Gaebel, F. Jelezko, and J. Wrachtrup, Science \textbf{320}, 1326
(2008).

\bibitem{mintert} F. Mintert and C. Wunderlich, Phys. Rev. Lett. \textbf{87}, 257904 (2001).

\bibitem{arbes} F. Arbes, M. Benzing, Th Gudjons, F. Kurth and G. Werth, Z.
Phys. D \textbf{31}, 27 (1994).

\bibitem{blatt1} J. Benhelm, G. Kirchmair, U. Rapol, T. Korber, C. F. Roos and R. Blatt,
Phys. Rev. A \textbf{75}, 032506 (2007).

\bibitem {deng} Z. J. Deng, M. Feng and K. L. Gao, Phys. Lett. A \textbf{344}, 97 (2005).

\bibitem {twam1} D. Mc Hugh and J. Twamley, Phys. Rev. A \textbf{71}, 012315 (2005).

\bibitem {KumarSelP} N. Sinha and T. S. Mahesh and K. V. Ramanathan and A. Kumar, J. Chem. Phys. \textbf{114}, 4415 (2001).

\bibitem{NielsenChuang} M. A. Nielsen and I. L. Chuang, {\it Quantum computation and quantum information},
Cambridge University Press, Cambridge (2001).

\bibitem{StolzeSuter} J. Stolze and D. Suter, {\it Quantum Computing: A Short Course from Theory to
Experiment}, Wiley-VCH, Berlin, 2nd edition (2008).

\bibitem{MicroRes} R. Narkowicz, D. Suter, and I. Niemeyer, Rev. Sci. Instrum. \textbf{79}, 084702 (2008).

\bibitem{bw} R. Blatt and D. Wineland, Nature (London) \textbf {453}, 1008 (1988).

\bibitem{bz} R. Blatt and P. Zoller, Euro. J. Phys. \textbf{9}, 250 (1988).

\bibitem{suterJCP} D. Suter and T. S. Mahesh, J. Chem. Phys. \textbf{128}, 052206 (2008).

\bibitem{Vandersypen} L. M. K. Vandersypen and I. L. Chuang, Rev. Mod. Phys.  \textbf{76}, 1037 (2004).

\bibitem{Pravia}
M. A. Pravia, N. Boulant, J. Emerson, A. Farid, E. M. Fortunato, T.
F. Havel, R. Martinez, and D. G. Cory, J. Chem. Phys.
 \textbf{119}, 9993 (2003).

\bibitem{Schulte}
T. Schulte-Herbr\"{u}ggen and O.W. S\"{o}rensen, Conc. Magn. Reson. \textbf{12}, 389 (2000).

\bibitem {lidar} D. A. Lidar, I. L. Chuang, and K. B. Whaley. Phys. Rev. Lett., \textbf{81}, 2594 (1998).

\bibitem{Schm} A. D. Cronin, J. Schmiedmayer, and D. E. Pritchard, arXiv 0712.3703v1. To appear in Rev. Mod. Phys.

\bibitem {sm} A. S{\o}rensen and K. M{\o}lmer, Phys. Rev. Lett. \textbf{82}, 1971 (1999).


\end{thebibliography}
\end{document}